\documentclass[letterpaper, 10 pt, conference]{ieeeconf}  

\IEEEoverridecommandlockouts                              

\usepackage{graphicx}
\usepackage{amsmath,amssymb,amsfonts}
\usepackage{cancel}
\usepackage[utf8]{inputenc}
\usepackage[T1]{fontenc}

\title{\LARGE \bf
Early Detection of Cognitive Impairment in Elderly using a Passive FPVS-EEG BCI and Machine Learning \\
-- Extended Version --}

\author{Tomasz M. Rutkowski$^{1,2,3,\ddag}$,  Stanisław Narębski$^2$, Mihoko Otake-Matsuura$^1$, and Tomasz Komendziński$^{2}$
\thanks{$^\ddag$Corresponding author.}
\thanks{$^{1}$Tomasz M. Rutkowski, and Mihoko Otake-Matsuura are with Cognitive Behavioral Assistive Technology Team (CBAT), RIKEN Center for Advanced Intelligence Project (RIKEN AIP), Tokyo, Japan,  {\tt\small tomasz.rutkowski@riken.jp}}%
\thanks{$^{2}$Tomasz M. Rutkowski, Stanisław Narębski, and Tomasz Komendziński are with the Department of Cognitive Science, Nicolaus Copernicus University, Toruń, Poland.}%
\thanks{$^{3}$Tomasz M. Rutkowski is with the Graduate School of Education, The University of Tokyo, Tokyo, Japan.}%
}

\begin{document}

\maketitle
\thispagestyle{empty}
\pagestyle{empty}

\begin{abstract}

Early dementia diagnosis requires biomarkers sensitive to both structural and functional brain changes. While structural neuroimaging biomarkers have progressed significantly, objective functional biomarkers of early cognitive decline remain a critical unmet need.  Current cognitive assessments often rely on behavioral responses, making them susceptible to factors like effort, practice effects, and educational background, thereby hindering early and accurate detection. This work introduces a novel approach, leveraging a lightweight convolutional neural network (CNN) to infer cognitive impairment levels directly from electroencephalography (EEG) data.  Critically, this method employs a passive fast periodic visual stimulation (FPVS) paradigm, eliminating the need for explicit behavioral responses or task comprehension from the participant. This passive approach provides an objective measure of working memory function, independent of confounding factors inherent in active cognitive tasks, and offers a promising new avenue for early and unbiased detection of cognitive decline.
\newline

\indent \textit{Clinical relevance}— This work establishes a potential neurophysiological biomarker for the early detection of cognitive impairment using a simplified fast periodic visual stimulation (FPVS) paradigm.
\end{abstract}

\section{INTRODUCTION}

This study investigates the potential of a passive fast periodic visual stimulation (FPVS)~\cite{fastball} brain-computer interface (BCI) to detect cognitive impairment (CI) and age-related changes, especially in the elderly. Early dementia may involve disruptions in the balance of excitatory and inhibitory neuronal activity, affecting sensory processing, including vision~\cite{cardin2018inhibitory}. This passive BCI approach offers a promising avenue for early detection and monitoring, as disruptions in this balance can lead to altered states of awareness. For example, some studies suggest that an overabundance of inhibitory activity might contribute to reduced awareness or even unconsciousness~\cite{cardin2018inhibitory}. 

Working memory (WM) relies on distinct oscillatory codes within the gamma band, with different types of information requiring different frequencies.  For instance, sensory-spatial WM maintenance has been linked to an alpha-gamma code, while sequential WM involves theta/gamma oscillations~\cite{hughes2008gamma,roux2014working}.  Given this established role of gamma oscillations in various WM tasks, and to mitigate the risk of muscle artifacts contaminating lower EEG frequencies~\cite{tomekJCSC2010}, this study focuses on higher gamma frequencies, specifically within the $30\sim125$~Hz range.  This analysis is performed on EEG recordings acquired at a $250$~Hz sampling frequency.

\section{METHODS}

\begin{figure}
\centerline{\includegraphics[width=\columnwidth]{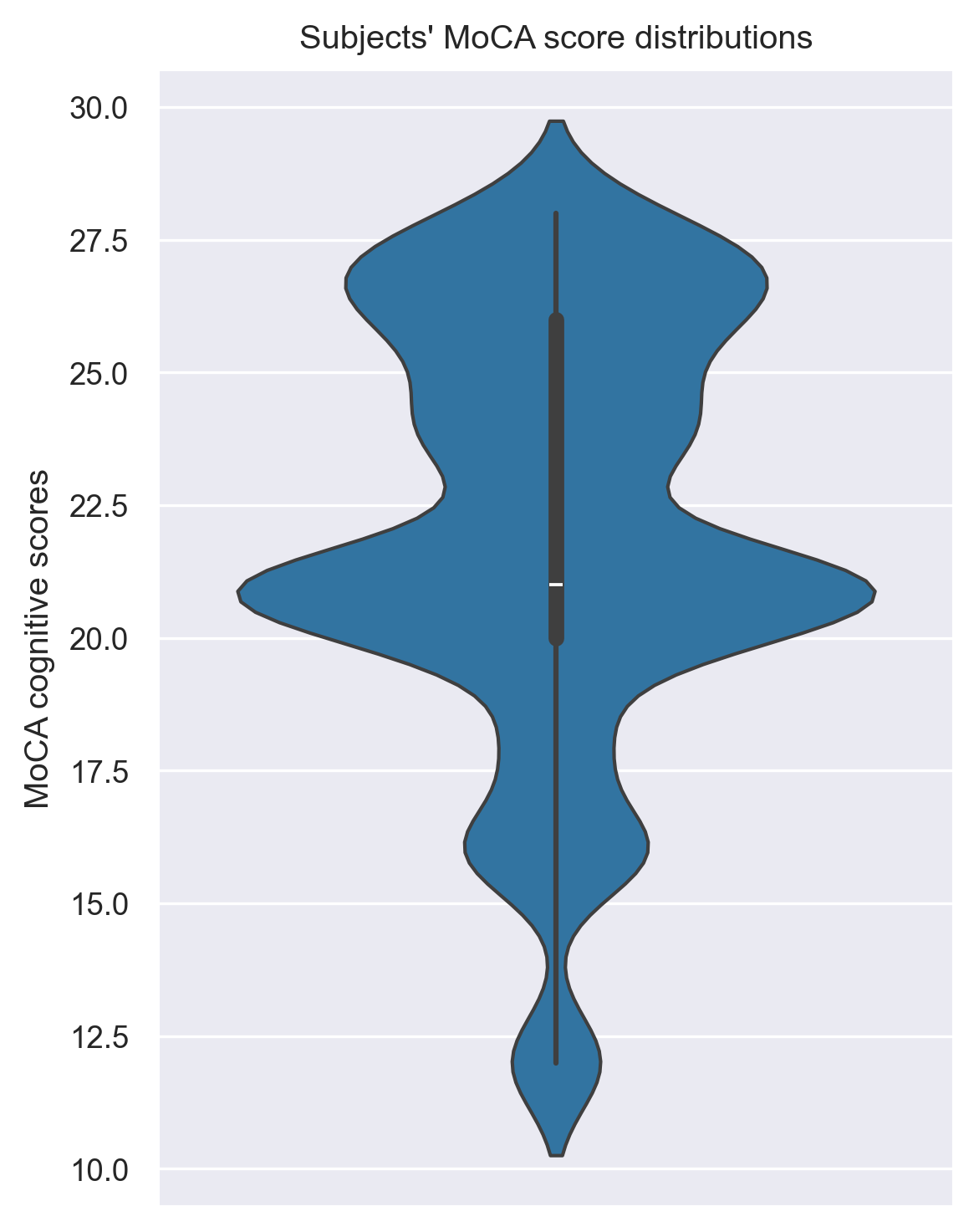}}
\caption{Distribution of MoCA scores among participants, showing median, quartiles, and range.}
\label{fig:moca}
\end{figure}

A feasibility study was conducted to evaluate the usability and viability of a passive visual BCI for assessing age-related CI.  Twenty-three elderly participants with varying levels of CI (see Figure~\ref{fig:moca}) took part in the study.  Participants performed a  visual BCI task, during which EEG data were recorded~\cite{tomekGBCI2024}. 

\begin{figure}
\centerline{\includegraphics[width=\columnwidth]{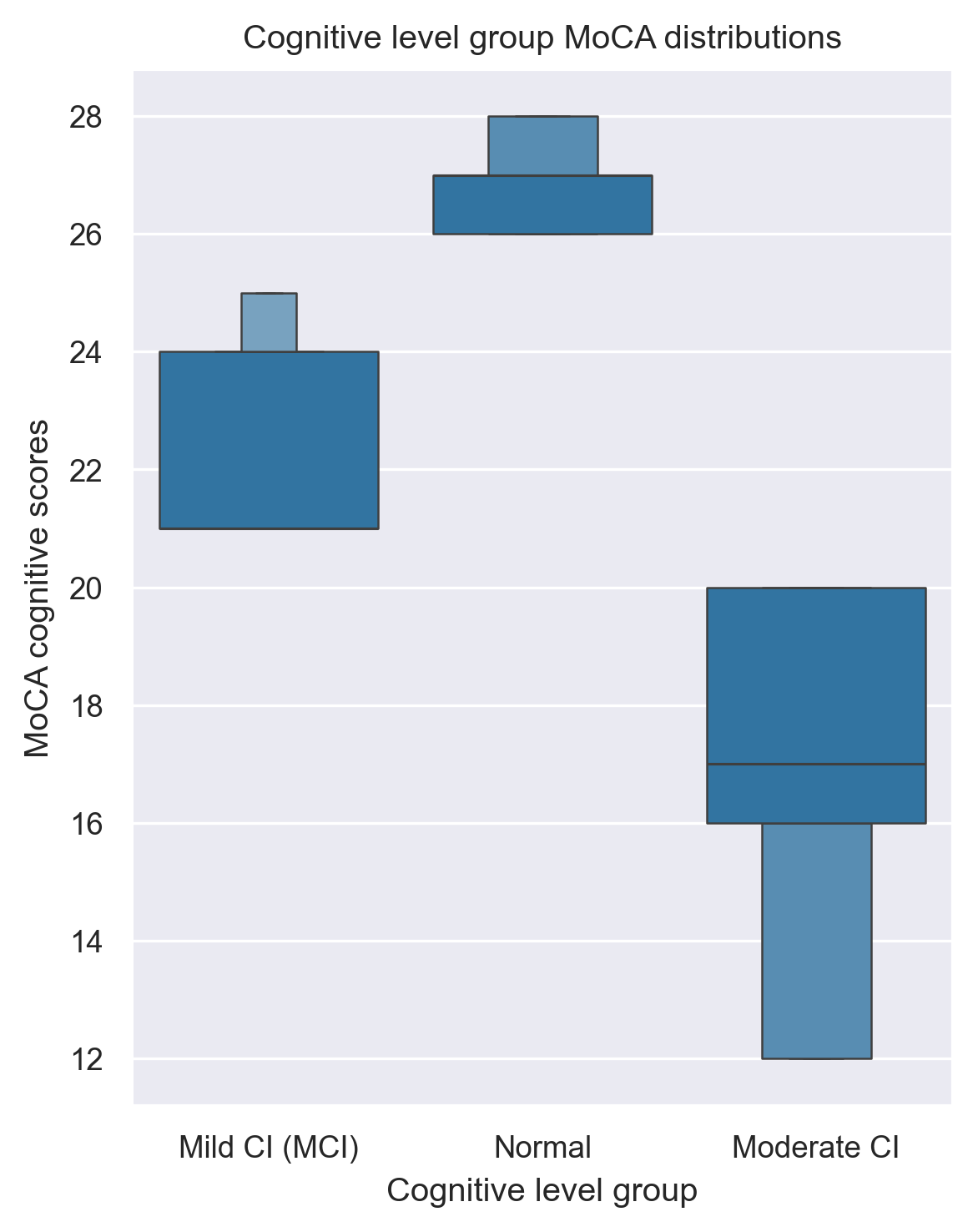}}
\caption{Distribution of participants across CI groups. These group labels were used for machine learning model training and subsequent prediction.} 
\label{fig:CI}
\end{figure}

This pilot study, approved by the Institute of Psychology NCU Ethical Committee and conducted in summer 2022 at Nicolaus Copernicus University (Toruń, Poland) in accordance with the Declaration of Helsinki, investigated a passive visual BCI paradigm using EEG.  Twenty-three female participants (mean age $70.5\pm 6.1$ years) were recruited.  CI levels were predicted using a lightweight convolutional neural network (CNN)~\cite{EEGNet} trained on data from a passive fast periodic visual stimulation (FPVS) paradigm using the so-called ``FastBall'' setting~\cite{fastball}.  This prediction task was framed as a three-class classification problem, corresponding to Montral Congnitve Assesment (MoCA)~\cite{moca2012} score ranges: normal ($26\sim 30$; $n = 7$), mild CI (MCI, $18\sim25$; $n = 13$), and moderate CI ($10\sim 17$; $n = 3$).  The distribution of CI groups is illustrated in Figure~\ref{fig:CI}.  Given the imbalanced class distribution, balanced accuracy was used as the evaluation metric in a ten-fold cross-validation scheme.  This passive BCI application focused on estimating CI based on visual stimulation, rather than generating control commands.

\subsection{FSVP Stimulus in ``FastBall'' Paradigm Setting}

Building on the recent adaptation of FPVS for passive, objective recognition memory measurement in Alzheimer's disease affected adults~\cite{fastball}, this project investigates its utility in predicting cognitive status (normal, MCI, and moderate CI) in elderly individuals.  Recognition memory, encompassing familiarity (a sense of prior exposure) and recollection (conscious retrieval of associated information), relies on brain regions (perirhinal cortex and hippocampus) known to be affected in early Alzheimer's disease~\cite{fastball}.  Utilizing the FPVS paradigm, standard stimuli are presented at $5$~Hz, with embedded oddball stimuli at a $1$~Hz equivalent rate (Figure~\ref{fig:experiment}).  This study aims to passively and objectively measure recognition performance in early dementia-related CI using FPVS, and to predict cognitive status from EEG data via machine learning.  Given that working memory decline is a hallmark of early dementia, and that such decline is associated with medial temporal lobe pathology~\cite{hughes2008gamma, roux2014working}, we hypothesize that cognitively impaired individuals will exhibit reduced recognition memory performance, reflected in altered FPVS responses within the gamma frequency band, compared to healthy older adults~\cite{fastball}.

\subsection{EEG Recording and Machine Learning Model}

\begin{figure}[t]
\centerline{\includegraphics[width=\columnwidth]{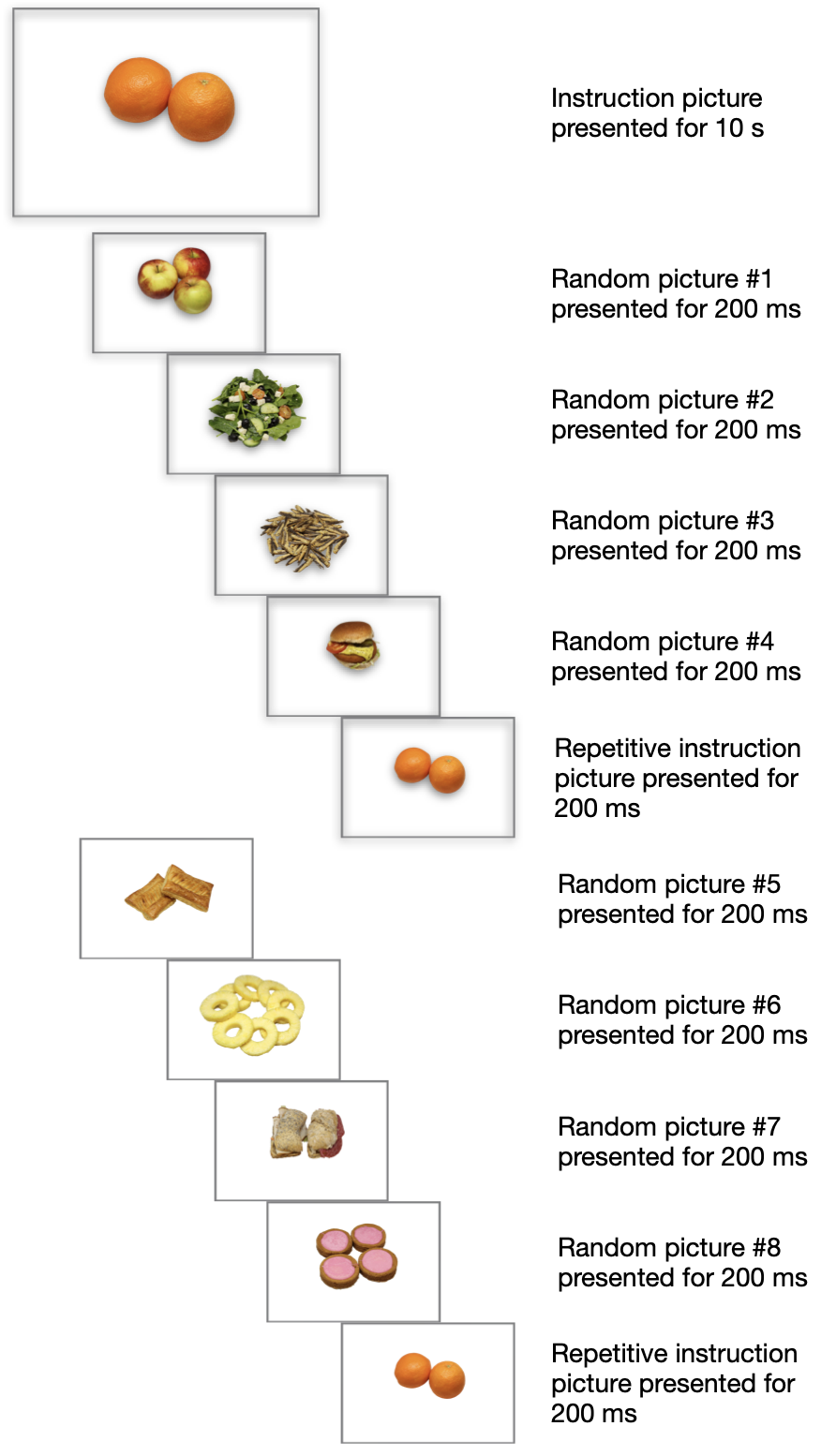}}
\caption{The FPVS image sequence employed in this study, featuring food photographs from the CROCUFID database~\cite{toet2019crocufid} and based on the ``FastBall'' paradigm~\cite{fastball}}
\label{fig:experiment}
\end{figure}

This pilot study introduces a novel approach using a portable and practical EEG system for passive assessment of cognitive function in elderly individuals.  EEG data were acquired using the Unicorn EEG headset (g.tec medical engineering, Austria), a wearable device previously validated against other EEG systems~\cite{tomekSCIS2022_1, 10.3389/fnhum.2023.1155194} and notable for its ease of use with dry electrodes.  Eight dry EEG channels ($Fz, C3, Cz, C4, Pz, PO7, Oz,$ and $PO8$) were recorded while participants passively viewed FPVS food image sequences (Figure~\ref{fig:experiment}).  This combination of a practical, dry-electrode EEG system with a passive viewing paradigm represents a significant step towards accessible and scalable cognitive assessment. 

Participants passively viewed an FPVS sequence, starting with a ten-second target image familiarization period followed by a two-minute "FastBall" presentation~\cite{fastball} with the target image occupying the same screen position (Figure~\ref{fig:experiment}). The total experimental duration for each participant was just two minutes, highlighting the efficiency of this paradigm.

EEG time-series data were digitized at 250 Hz. Minimal preprocessing, consisting only of bandpass filtering in the ranges of $0.5\sim4$Hz ($\delta-$wave), $4\sim8$Hz ($\theta-$wave), $8\sim14$Hz ($\alpha-$wave), $14\sim30$Hz ($\beta-$wave), $30\sim125$Hz (broad $gamma-$wave), $0.5\sim125$Hz (full-band EEG)~\cite{MNE} and 50 Hz notch filtering~\cite{MNE}, were applied.  One-second epochs corresponding to a single ``FastBall'' sequence, aligned with the stimulus presentation paradigm, were extracted.  Critically, an end-to-end machine learning approach~\cite{braindecode2017, EEGNet} was adopted, eliminating the need for extensive feature engineering.  Specifically, the EEGNetv4 architecture~\cite{EEGNet}, implemented using~\cite{braindecode}, was employed with $64-$length input convolutional filters, the $AdamW$ optimizer, a mini-batch size of $8$, an initial learning rate of $0.005$, a maximum of $200$ epochs, and ten-fold randomized cross-validation~\cite{scikit-learn}. This end-to-end training paradigm enabled the simultaneous optimization of implicit denoising and multi-class classification~\cite{braindecode2017, EEGNet}, directly mapping raw (minimally preprocessed) EEG data to cognitive status predictions.

\section{RESULTS}

\begin{figure*}[t]
\centerline{\includegraphics[width=\textwidth]{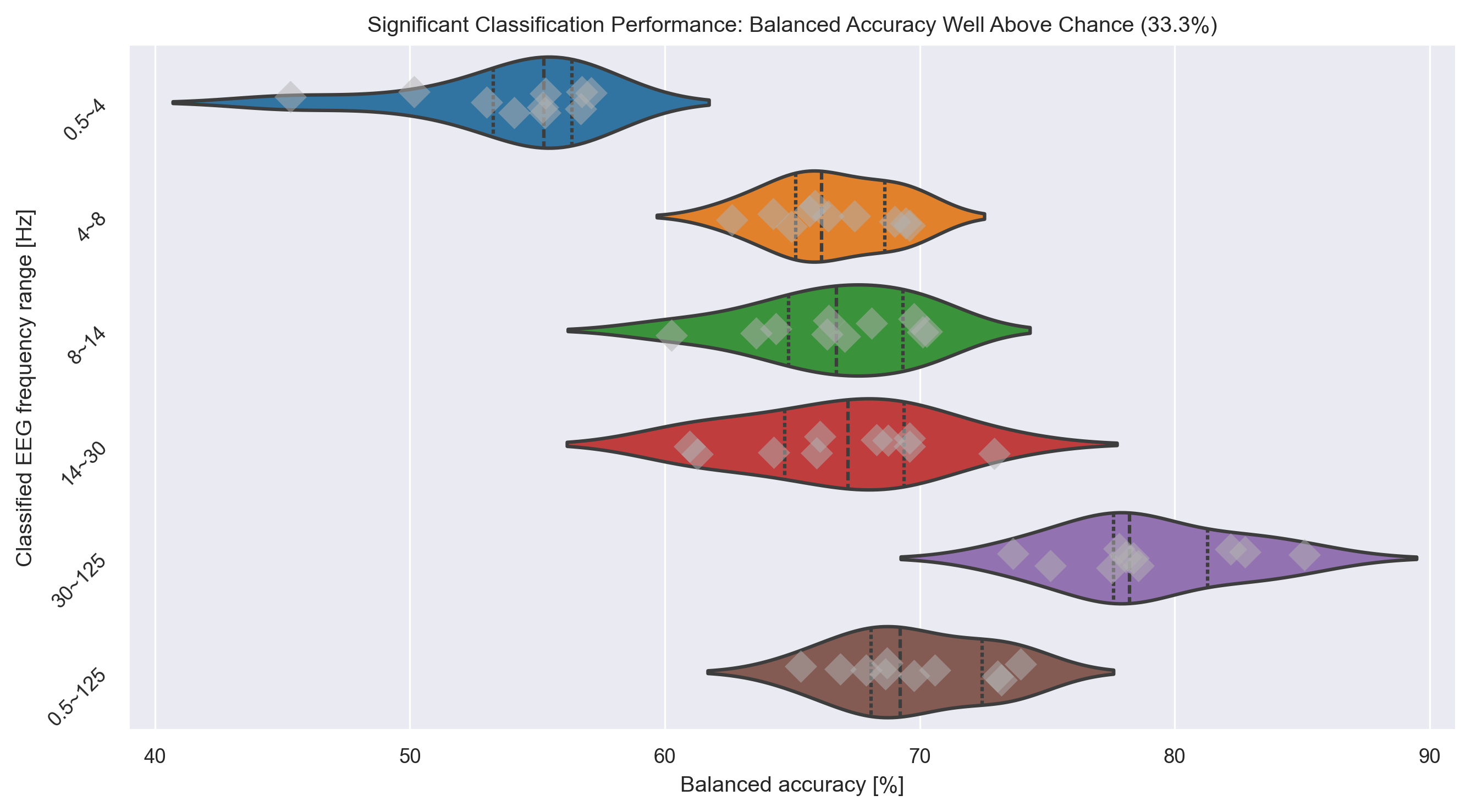}}
\caption{Ten-fold cross-validation balanced accuracies across EEG frequency bands, with overlaid median, quartiles, and range. Chance level: $33.3\%.$}
\label{fig:accRESULTS}
\end{figure*}

A pilot study with 23 elderly participants investigated activity in various EEG frequency bands during a passive visual BCI task using FSVP.  
Ten-fold cross-validation was employed due to the limited sample size, and resulting balanced accuracies for various EEG frequency bands were: $55.3\%$ ($0.5\sim4$~Hz, $\delta$), $66.2\%$ ($4\sim8$~Hz, $\theta$), $66.7\%$ ($8\sim14$~Hz, $\alpha$), $67.2\%$ ($14\sim30$~Hz, $\beta$), $78.2\%$ ($30\sim125$~Hz, broad $\gamma$), and $69.2\%$ ($0.5\sim125$~Hz, $full-band$). Figure~\ref{fig:accRESULTS} presents the results as distribution plots.

All results were significantly above chance ($33.3\%$; $p\ll0.01$, rank-sum test~\cite{scipy}).  The $\delta-$band ($0.5\sim4$~Hz) showed significantly lower accuracy than all other bands ($p < 0.01$). Critically, the broad $\gamma-$band ($30\sim125$~Hz) yielded significantly higher accuracy than all other bands ($p\ll0.01$). 
The broad $\gamma-$band ($30\sim125$~Hz) shows particular promise for CI estimation within this passive visual FSVP BCI paradigm, potentially reflecting dementia-related changes in neural activity~\cite{hughes2008gamma, roux2014working}

\section{CONCLUSIONS}

This pilot study establishes the feasibility of a novel passive FSVP-BCI approach for tracking CI in elderly individuals, offering a promising avenue for early detection and intervention in dementia-related decline.  While these initial findings are encouraging, further development of the paradigm, including optimized visual presentation and robust signal processing, is warranted.  Future research will involve a larger-scale study to refine and validate this non-invasive method for monitoring cognitive wellbeing and addressing the growing challenge of dementia.

\section*{AUTHOR CONTRIBUTIONS}

TMR, SN, and TK conceived the FSVP paradigm for CI prediction. TMR developed the end-to-end convolutional neural network approach. SN and TK performed the experiments and recruited participants. TMR designed and implemented the experimental procedures and data analysis. TMR analyzed the data; SN, TK, and MOM interpreted the results. TMR wrote the manuscript.

\section*{FUNDING}

MOM and TMR were partly supported by the Japan Science and Technology Agency AIP Trilateral  AI  Research  Grant  No.~JPMJCR20G1 from the Japan Science and Technology Agency. TMR received support from Nicolaus Copernicus University in Torun, Poland, through the 2022 and 2024 Mobility Grants. TK received support from Nicolaus Copernicus University's Emerging Field projects ``Cognition and Language 2022'' and ``Culture, Development \& Wellbeing'' in Torun, Poland.

\section*{ACKNOWLEDGMENT}

The authors gratefully acknowledge the senior volunteers from Kamienica Inicjatyw, Toruń, Poland, for their invaluable participation in this project. Their contributions were essential to the advancement of the EEG-based neuro-biomarker research presented herein.



\addtolength{\textheight}{-12cm}   

\end{document}